\documentclass[ijoc,sglanonrev]{informs4}
\usepackage{eqndefns-left} % For checking the display equation width and equation environment definitions %
\RequirePackage{tgtermes}
\RequirePackage{newtxtext}
\RequirePackage{newtxmath}
\RequirePackage{bm}
\RequirePackage{endnotes}

\OneAndAHalfSpacedXII % Current default line spacing
\usepackage{algorithm}
\usepackage{algpseudocode}
\usepackage{tikz}
\usetikzlibrary{positioning,matrix}
\usetikzlibrary{arrows.meta,positioning,calc}

\usepackage{natbib}
 \bibpunct[, ]{(}{)}{,}{a}{}{,}%
 \def\bibfont{\small}%
\EquationsNumberedThrough    % Default: (1), (2), ...
\TheoremsNumberedThrough     % Preferred (Theorem 1, Lemma 1, Theorem 2)
\ECRepeatTheorems  %  

\MANUSCRIPTNO{IJOC-0001-2024.00}

\begin{document}
%%%%%%%%%%%%%%%%

%\setcounter{page}{1}
%\DOI{10.1287/xxxx.0000.0000}%

\RUNAUTHOR{Wang}

% \RUNTITLE{Predictive Maintenance in Manufacturing}
\RUNTITLE{Conditional Cauchy--Schwarz Divergence for Time-Series Analysis}

% \TITLE{Optimal Resource Allocation in Humanitarian Logistics: A Stochastic Programming Approach}
\TITLE{Conditional Cauchy--Schwarz Divergence for Time-Series Analysis: Kernelized Estimation and Applications in Clustering and Fraud Detection}

\ARTICLEAUTHORS{%
%\AUTHOR{John Doe,\textsuperscript{a} Jane Smith,\textsuperscript{b}}
%\AFF{\textsuperscript{a}Department of Industrial Engineering, University of XYZ, \EMAIL{john.doe@xyz.edu; \textsuperscript{b}Department of Computer Science, University of ABC, \EMAIL{jane.smith@abc.edu}} 
\AUTHOR{Jiayi Wang}
\AFF{Imperial College Business School, Imperial College London, Exhibition Road, London, SW7 2AZ, United Kingdom, \EMAIL{wangjy030409@gmail.com}}}

\ABSTRACT{%
% Enter your abstract
We study the conditional Cauchy--Schwarz divergence (C-CSD) as a generic, symmetric, density-free measure of discrepancy for time-series analysis. We derive a practical kernel-based estimator using radial basis function (RBF) kernels on both the condition and output spaces, together with several numerical stabilizations, including a symmetric logarithmic form with an epsilon ridge and a robust bandwidth selection rule based on the interquartile range (IQR). Median-heuristic bandwidths are applied to the window vectors, and effective-rank filtering is used to avoid degenerate kernels. We demonstrate the framework in two applications. In time-series clustering, conditioning on the time index and comparing scalar series values yield a pairwise C-CSD dissimilarity. Each series is standardized within its own data split to prevent information leakage. The time-kernel bandwidth and the value-kernel bandwidth are selected on the training split, after which precomputed-distance k-medoids clustering is performed on the test split, and normalized mutual information (NMI) is reported over multiple random seeds. In fraud detection, conditioning on sliding transaction windows and comparing the magnitude of value changes with categorical and merchant change indicators, we score each query window by contrasting a global normal-reference mixture against a same-account local-history mixture with recency decay and change-flag weighting. Account-level decisions are obtained by aggregating window scores using the maximum value. Experiments on four benchmark time-series datasets and a transactional fraud-detection dataset demonstrate that the C-CSD estimator is numerically stable and supports effective clustering and anomaly detection under a strictly leak-proof evaluation protocol.
}%

%Supplemental Material:
%Data Ethics & Reproducibility Note:

%\KEYWORDS{Stochastic programming, Decision support,Uncertainty, Disaster response, Optimization}

%\HISTORY{Received: Month DD, YYYY; Accepted: Month DD, YYYY; Published Online: Month DD, YYYY}

\maketitle
%%%%%%%%%%%%%%%%%%%%%%%%%%%%%%%%%%%%%%%%%%%%%%%%%%%%%%%%%%%%%%%%%%%%%%

% Text of your paper here

\section{Introduction}\label{sec:Intro}

Measuring discrepancies between time-series is a central problem in unsupervised learning, with direct relevance to clustering, anomaly detection, and sequential decision making. In practice, the dominant approach relies on distance-based similarity measures that operate directly on observed sequences. Among them, Dynamic Time Warping (DTW) remains the most widely used method due to its ability to accommodate local temporal misalignment \citep{berndt1994dtw,sakoe1978}. DTW and its constrained variants have been successfully applied to time-series clustering and retrieval, typically in combination with medoid-based backends \citep{petitjean2011dba}. However, DTW-based prototype learning is inherently nonsmooth and difficult to optimize, which complicates centroid estimation and limits its integration with distributional or kernel-based methods \citep{schultz2018dtw}. More broadly, purely geometric distances focus on alignment costs and do not explicitly characterize regime-dependent or distributional discrepancies that are often critical in complex temporal data.

An alternative line of work compares time series at the distributional level. Divergence-based criteria aim to quantify discrepancies between probability laws rather than individual trajectories, thereby offering robustness to noise and local variability. Classical divergences such as the Kullback--Leibler divergence are sensitive but typically rely on density estimation or density ratios, which can be unstable in finite-sample regimes. Kernel-based integral probability metrics, most notably the maximum mean discrepancy (MMD), avoid explicit density estimation and admit simple empirical estimators \citep{gretton2012mmd}. Extensions based on kernel mean embeddings further enable comparisons of conditional distributions \citep{song2009conditional,muandet2017kme}. Nevertheless, in the conditional setting these criteria are not faithful: a zero discrepancy does not, in general, imply equality of conditional distributions. As a result, they may fail to distinguish genuinely different conditional laws that agree in their embedded moments.

The conditional Cauchy--Schwarz divergence (C-CSD) introduced by Yu et al.~\citep{yu2023conditional} addresses this limitation by providing a symmetric, density-free, and faithful measure of discrepancy between conditional distributions. Building on the Parzen-window interpretation of the Cauchy--Schwarz divergence \citep{jenssen2006cs}, C-CSD admits a kernel-based estimator that reduces conditional density inner products to Gram-matrix traces, avoiding density-ratio estimation while preserving faithfulness. Despite these theoretical advantages, the practical use of C-CSD in time-series analysis remains limited, particularly in application-driven settings that require strict evaluation protocols and numerical robustness. In this work, we study a stabilized kernelized estimator of C-CSD and demonstrate its effectiveness in two representative tasks: time-series clustering and transactional fraud detection. Our focus is not on proposing a new divergence, but on showing how C-CSD can be instantiated, stabilized, and deployed as a practical discrepancy measure under leak-proof experimental protocols.

\section{Related Work}
\subsection{CS Divergence: Foundations and Uses}
The Cauchy--Schwarz (CS) divergence is a symmetric and density-ratio-free measure whose Parzen--window interpretation connects naturally to kernel and spectral criteria~\citep{jenssen2006cs}. Kernel two-sample statistics such as the maximum mean discrepancy (MMD) further position CS-style kernel density and Gram-based estimators as numerically stable alternatives to the Kullback--Leibler divergence in finite-sample regimes~\citep{gretton2012mmd}. Because of these properties, the CS divergence provides a practical foundation for kernel formulations of information measures in both clustering and representation learning.

Following the conditional CS divergence introduced by Yu et~al.~\citep{yu2023conditional}, we adopt the same sample estimator, expressed as a kernel density or Gram trace form that aggregates within- and cross-conditional similarities using row-normalized weights and logarithmic aggregation. Our work does not propose a new estimator. The differences lie in two aspects. First, we design task-specific conditioning and reference sets for time-series problems: in clustering, conditioning on the time index; in fraud detection, using window-level conditioning that contrasts a global reference library of normal samples with a local, past-only account history and account-level maximum aggregation. Second, we introduce numerical and operational stabilizations, including a symmetric logarithmic form with an epsilon ridge, effective-rank filtering, and a leak-proof protocol in which bandwidths and thresholds are fixed from validation. These choices preserve the estimator in~\citep{yu2023conditional} while improving robustness and deployability in our settings (see Sec.~4.1 and Sec.~4.2).

\subsection{Time-Series Clustering}
Time-series clustering is commonly organized as a pipeline of representation, pairwise dissimilarity, and a clustering algorithm. Distance-based approaches measure similarity directly on the raw series. Dynamic Time Warping remains the standard elastic distance because it tolerates local time shifts and dilations; we follow prior work and use k-medoids on a precomputed distance matrix as a typical clustering choice~\citep{sakoe1978,berndt1994dtw}. A practical difficulty is how to define centroids under alignment. DTW-consistent averaging addresses this by iteratively refining a prototype that respects alignment paths and improves centroid-based clustering in practice~\citep{petitjean2011dba}.

Edit-based distances handle missing data, noise, and unequal lengths by allowing insertions, deletions, and substitutions. Representative variants include the Longest Common Subsequence, Edit Distance with Real Penalty, and Edit Distance on Real sequence~\citep{vlachos2002lcss,chen2004erp,chen2005edr}. Compared with elastic distances, these measures are more forgiving to gaps and local mismatches but introduce thresholds or penalties that are task dependent. All of them can be used to form pairwise dissimilarity matrices for k-medoids.

Shape-based methods target phase-invariant comparison and learn representative shapes as cluster centers. The k-Shape algorithm uses a z-normalized, phase-invariant correlation distance together with an efficient centroid update and often serves as a strong domain-independent baseline~\citep{paparrizos2015kshape}. Because it avoids explicit dynamic programming over alignment paths, it scales well to long sequences while preserving interpretability through prototypical shapes.

Kernel and differentiable formulations embed alignment into a positive-definite similarity or replace the hard minimum over alignments with a smooth objective. The global alignment kernel sums contributions over alignment paths and enables spectral or kernel clustering~\citep{cuturi2011gak}. Soft-DTW provides a differentiable loss that exposes gradients of the alignment objective and is widely used upstream of clustering to learn representations~\citep{cuturi2017softdtw}. In contrast, our study treats dissimilarity as a conditional divergence between distributions under a shared temporal reference. The conditional Cauchy–Schwarz estimator constructs Gram matrices with radial basis function kernels on the condition space and the value space and aggregates them in a numerically stable logarithmic form, yielding a symmetric and density-free dissimilarity that avoids explicit path search and integrates naturally with k-medoids~\citep{yu2023conditional}.

Feature and encoding approaches form an alternative line by mapping each series to a vector of hand-crafted or transform-based features and then clustering in that space. Toolkits such as \texttt{tsfresh} and catch22 provide broad libraries of interpretable statistics; symbolic and transform representations such as Symbolic Aggregate approXimation and multiscale transforms, including Fourier, wavelet, and scattering transforms, reduce dimensionality while exposing structure~\citep{christ2018tsfresh,lubba2019catch22,lin2007sax,bruna2013scattering}. These methods scale and are interpretable but require feature selection or domain tuning. Our approach differs in that it computes dissimilarity directly on window vectors with data-driven kernel bandwidths and does not rely on an external feature vocabulary.

Representation learning methods first learn embeddings and then apply a conventional clustering step. Recent self-supervised techniques such as hierarchical contrastive learning and temporal–context contrasting produce robust embeddings across scales and augmentations and can be effective for multivariate data at scale~\citep{yue2022ts2vec,eldele2021tstcc}. Model-based clustering with graphical dependencies, such as Toeplitz inverse covariance-based clustering, jointly segments and clusters multivariate series~\citep{hallac2017ticc}. These approaches require dedicated training and careful validation to avoid leakage. By comparison, the conditional Cauchy–Schwarz dissimilarity is training-free beyond bandwidth selection and emphasizes stable estimation under strict split protocols. We therefore adopt DTW with k-medoids as the elastic baseline and position the conditional Cauchy–Schwarz dissimilarity as a kernel-style alternative focused on agreement of local dynamics under a shared temporal reference~\citep{tsclust_survey_2024}.

\subsection{Fraud Detection on Transaction Time Series}
Fraud detection on transactional time series spans supervised, semi-supervised, and unsupervised settings, with transaction-level scores often aggregated to the account level for decisions. In supervised sequence modeling, feature-based learners and deep temporal models are trained on labeled transactions under heavy class imbalance~\citep{he2009imbalance}. Sequence classification with long short-term memory networks exploits temporal dependencies and improves over static learners on credit-card data~\citep{jurgovsky2018sequence}, while Transformer variants further enhance long-range dependency modeling for anomaly and fraud detection~\citep{zhou2021anomalytransformer}. Operating points are typically selected on a separate validation split rather than fixed percentiles, and probability calibration with rebalancing is recommended for imbalanced regimes~\citep{dalpozzo2015calibrating}. Practical deployments also report concept drift and delayed feedback in streams, motivating periodic retraining and adaptive thresholds~\citep{dalpozzo2015drift,dalpozzo2014lessons}.

Graph and relational approaches model customers, merchants, devices, and transactions as a heterogeneous network so that risk propagates through neighborhoods and relational structures. Recent graph neural networks introduce augmentation and robust message passing to address sparsity and heterophily; Dual-Augment GNN is a representative design for transaction fraud~\citep{li2022dualaugment}. Surveys summarize progress in deep graph anomaly detection and highlight challenges for evolving graphs and scalability~\citep{ma2021graphanom}.

Unsupervised and one-class methods avoid reliance on fraud labels. Density- and distance-based detectors such as local outlier factor and Isolation Forest provide generic scoring for rare deviations~\citep{breunig2000lof,liu2008iforest}. One-class support vector machines estimate the support of the normal distribution and are widely used when only normal data are available~\citep{scholkopf2001ocsvm}. Reconstruction- or prediction-based approaches detect anomalies by large reconstruction or forecasting errors, for example with autoencoders and LSTM encoder–decoder models~\citep{malhotra2016encdec}. Comparative evaluations on time-series benchmarks emphasize protocol design (split hygiene, leakage control, and metric choice) and show that algorithm families trade off sensitivity and robustness depending on regime and preprocessing~\citep{wenig2022vldb}. For extremely imbalanced data, average precision or area under the precision–recall curve is preferred over ROC-style metrics, and thresholds should be tuned on held-out validation sets~\citep{he2009imbalance,dalpozzo2015calibrating}.

Our study follows the unsupervised line on BankSim-style transactional data. We construct a global normal reference library from disjoint accounts and, for each query window, contrast it with a same-account local history under a shared temporal reference~\citep{banksim2014,banksimKaggle}. The score is a conditional Cauchy–Schwarz divergence computed with kernel Gram matrices on the condition space and the value-change space; this yields a symmetric, density-free dissimilarity without explicit path search, and it integrates naturally with account-level aggregation by the maximum over windows~\citep{yu2023conditional}. Bandwidths and summary statistics are estimated on the normal-only library, the operating threshold is fixed on validation, and test reporting is strictly leak-proof with account disjointness and temporal ordering~\citep{banksim2014}.

\section{Methodology}\label{sec:Method}
Section 3 introduces the conditional Cauchy--Schwarz divergence and the kernelized estimator used throughout this work. 
We first recall the definition and key properties of the conditional Cauchy--Schwarz divergence.
We then derive a finite-sample kernel/Gram estimator based on Parzen windowing and discuss numerical regularization strategies.
Finally, we summarize the complete algorithmic pipeline that underlies both the clustering and fraud-detection applications.

\subsection{Conditional Cauchy--Schwarz Divergence}\label{subsec:ccs}

Let $\mathcal{X}$ be the \emph{condition} space and $\mathcal{Y}$ be the \emph{output} space.
We compare two families of conditional densities $p(\cdot\mid x)$ and $q(\cdot\mid x)$ on $\mathcal{Y}$, indexed by $x\in\mathcal{X}$,
with respect to a reference measure $r$ on $\mathcal{X}$.
Write the $L^2(\mathcal{Y})$ inner product and norm as
$\langle f,g\rangle \!=\! \int_{\mathcal{Y}} f(y)g(y)\,dy$ and $\|f\|_2^2 \!=\! \langle f,f\rangle$.
Throughout, we use positive–definite kernels on the condition and output domains:
$K_\tau:\mathcal{X}\!\times\!\mathcal{X}\!\to\!\mathbb{R}$ and $L_\sigma:\mathcal{Y}\!\times\!\mathcal{Y}\!\to\!\mathbb{R}$,
with bandwidths (hyper–parameters) $\tau,\sigma>0$.
The unconditional Cauchy--Schwarz (CS) divergence admits a classical Parzen–window interpretation
where density inner products reduce to kernel Gram sums~\citep{jenssen2006cs};
we will leverage this viewpoint in the conditional setting in \S\;3.3.
For context, kernel two–sample/IPM criteria (e.g., MMD) and kernel mean embeddings
provide alternative tools for (conditional) distribution comparison~\citep{gretton2012mmd,muandet2017kme,song2009conditional},
but differ in faithfulness and estimator structure (see below).
We follow the definition and estimator paradigm of conditional CS divergences in~\citep{yu2023conditional},
while deriving an explicit kernel/Gram form suitable for finite–sample computation (cf.\ \S\;3.3).

Given conditional densities $p(\cdot\mid x)$ and $q(\cdot\mid x)$ and a reference distribution $r$ on $\mathcal{X}$, the conditional Cauchy--Schwarz (CS) divergence is defined as
\begin{equation}
\label{eq:ccsd-def}
D_{\mathrm{CS}}(p\Vert q \mid r)
=
-\log
\frac{\mathbb{E}_{x\sim r}\!\big[\langle p(\cdot\mid x),\,q(\cdot\mid x)\rangle\big]}
{\sqrt{\mathbb{E}_{x\sim r}\!\big[\|p(\cdot\mid x)\|_{2}^{2}\big]\,
       \mathbb{E}_{x\sim r}\!\big[\|q(\cdot\mid x)\|_{2}^{2}\big]}}\;,
\end{equation}
where $\langle f,g\rangle=\int f(y)g(y)\,dy$ and $\|f\|_{2}^{2}=\int f(y)^{2}\,dy$ denote the standard $L_{2}$ inner product and norm on the output space $\mathcal{Y}$.

To clarify the inequality that underlies this formulation, recall the Cauchy--Schwarz inequality in the $L_{2}$ space over the product measure $r(x)\,dx\otimes dy$:
\begin{equation}
\label{eq:cs-XY}
\Big|\!\iint f(x,y)\,g(x,y)\,r(x)\,dx\,dy\Big|
\le
\Big(\!\iint |f(x,y)|^{2}r(x)\,dx\,dy\Big)^{\!\frac12}
\Big(\!\iint |g(x,y)|^{2}r(x)\,dx\,dy\Big)^{\!\frac12},
\end{equation}
with equality if and only if $f=\lambda g$ almost everywhere for some scalar $\lambda$.
Substituting $f(x,y)=p(y\mid x)$ and $g(x,y)=q(y\mid x)$ into \eqref{eq:cs-XY} gives
\[
\mathbb{E}_{x\sim r}\!\big[\langle p(\cdot\mid x),\,q(\cdot\mid x)\rangle\big]
\le
\sqrt{\mathbb{E}_{x\sim r}\!\big[\|p(\cdot\mid x)\|_{2}^{2}\big]\,
      \mathbb{E}_{x\sim r}\!\big[\|q(\cdot\mid x)\|_{2}^{2}\big]},
\]
which immediately implies $D_{\mathrm{CS}}(p\Vert q\mid r)\ge0$.  
Equality in \eqref{eq:cs-XY} requires $p(\cdot\mid x)=\lambda(x)q(\cdot\mid x)$ for $r$-almost all $x$; because both are normalized conditional densities, $\lambda(x)\equiv1$, hence $p(\cdot\mid x)=q(\cdot\mid x)$ almost surely.  Therefore $D_{\mathrm{CS}}(p\Vert q\mid r)=0$ if and only if the two conditional distributions coincide under $r$, which ensures non-negativity, symmetry, and faithfulness of the divergence~\citep{yu2023conditional}.  

When $r$ degenerates to a point mass, \eqref{eq:ccsd-def} reduces to the standard (unconditional) CS divergence introduced by Jenssen et al.~\citep{jenssen2006cs}.  The Parzen-window interpretation of this form further motivates a kernel or Gram-matrix estimator for the conditional case, obtained by replacing inner products over $y$ with kernel evaluations; this practical estimator and its numerical safeguards are described in~\S3.3.

\subsection{Kernelized Estimator and Derivation}
We derive a computable estimator for the conditional CS divergence in \eqref{eq:ccsd-def}
using Parzen windows and Gram matrices, following the unconditional CS view in~\citep{jenssen2006cs}
and the conditional CS paradigm of~\citep{yu2023conditional}.

Let $S_p=\{(x_i^p,y_i^p)\}_{i=1}^{n_p}\!\sim p$ and $S_q=\{(x_j^q,y_j^q)\}_{j=1}^{n_q}\!\sim q$.
Let $K_\tau$ and $L_\sigma$ be positive–definite kernels on $\mathcal X$ and $\mathcal Y$ with bandwidths $\tau,\sigma>0$.
For a set of reference points $\{x_\ell\}_{\ell=1}^{n_r}$ that approximate the expectation over $x\!\sim\!r$,
define normalized condition–kernel weights
\[
\alpha_i(x)=\frac{K_\tau(x,x_i^p)}{\sum_{i'}K_\tau(x,x_{i'}^p)},
\qquad
\beta_j(x)=\frac{K_\tau(x,x_j^q)}{\sum_{j'}K_\tau(x,x_{j'}^q)}.
\]
The conditional KDEs take Nadaraya--Watson form~\citep{nadaraya1964estimating,watson1964smooth}:

\[
\widehat p(y\mid x)=\sum_{i=1}^{n_p}\alpha_i(x)\,L_\sigma(y,y_i^p),
\qquad
\widehat q(y\mid x)=\sum_{j=1}^{n_q}\beta_j(x)\,L_\sigma(y,y_j^q).
\]

In practice we take $\{x_\ell\}$ as either (i) the union $\{x_i^p\}\cup\{x_j^q\}$,
or (ii) a task-agnostic grid on $\mathcal X$. Both choices lead to the same estimator
(up to the numerical values of $A,B$) and do not alter the methodology.

Let
\[
(L_{pq})_{ij}=\int_{\mathcal Y} L_\sigma(y,y_i^p)L_\sigma(y,y_j^q)\,dy,\quad
(L_{pp})_{ii'}=\int_{\mathcal Y} L_\sigma(y,y_i^p)L_\sigma(y,y_{i'}^p)\,dy,\ \text{etc.}
\]
For Gaussian $L_\sigma$, the integral equals a Gaussian with a bandwidth rescaling times a positive constant
(convolution closure); constants cancel in the CS ratio, so we keep the notation $L_{pq},L_{pp},L_{qq}$ for brevity~\citep{jenssen2006cs}.
With $A_{\ell i}=\alpha_i(x_\ell)$ and $B_{\ell j}=\beta_j(x_\ell)$, and writing
$\boldsymbol a_\ell$ and $\boldsymbol b_\ell$ for the $\ell$-th rows of $A,B$, we obtain
\[
\widehat I_{pq}
=\frac{1}{n_r}\sum_{\ell=1}^{n_r}\boldsymbol a_\ell^{\!\top} L_{pq}\,\boldsymbol b_\ell
=\frac{1}{n_r}\,\mathrm{tr}\!\big(L_{pq}\,A^\top B\big),
\]
\[
\widehat I_{pp}=\frac{1}{n_r}\,\mathrm{tr}\!\big(L_{pp}\,A^\top A\big),\qquad
\widehat I_{qq}=\frac{1}{n_r}\,\mathrm{tr}\!\big(L_{qq}\,B^\top B\big).
\]
Note that $A\!\in\!\mathbb{R}^{n_r\times n_p}$ and $B\!\in\!\mathbb{R}^{n_r\times n_q}$, while
$L_{pq}\!\in\!\mathbb{R}^{n_p\times n_q}$, $L_{pp}\!\in\!\mathbb{R}^{n_p\times n_p}$ and
$L_{qq}\!\in\!\mathbb{R}^{n_q\times n_q}$; hence $\mathrm{tr}(L_{pq}A^\top B)$,
$\mathrm{tr}(L_{pp}A^\top A)$ and $\mathrm{tr}(L_{qq}B^\top B)$ are well-defined.
Moreover $A,B$ are row-stochastic (each row sums to $1$ by construction).

Plugging the above into \eqref{eq:ccsd-def} yields the computable kernel/Gram estimator
\begin{equation}
\label{eq:ccsd-est}
\widehat D_{\mathrm{CS}}
= -\log\!\left(
\frac{\mathrm{tr}\!\big(L_{pq}\,A^\top B\big)}
{\sqrt{\big(\mathrm{tr}(L_{pp}\,A^\top A)+\varepsilon\big)\,\big(\mathrm{tr}(L_{qq}\,B^\top B)+\varepsilon\big)}}
\right),
\end{equation}
where $\varepsilon>0$ ensures numerical stability when Gram terms are small.
If $p=q$ and the same sample is used on both sides, then $A=B$ and $L_{pq}=L_{pp}=L_{qq}$,
so the numerator equals the denominator and $\widehat D_{\mathrm{CS}}=0$, consistent with \S\;3.2.
Equation~(3) is algebraically consistent with the conditional Cauchy--Schwarz divergence estimator introduced in~\citep{yu2023conditional}, 
and only differs in finite-sample stabilizations used throughout our experiments. 
Specifically, we add a small ridge term~$\varepsilon>0$ to avoid numerical instability when Gram blocks are nearly singular; 
we adopt a symmetric logarithmic form 
$-\tfrac{1}{2}\,[\log(I_{pp}{+}\varepsilon)+\log(I_{qq}{+}\varepsilon)-2\log(I_{pq}{+}\varepsilon)]$
to maintain symmetry in $(p,q)$ and prevent negative rounding artifacts; 
and we implement the estimator explicitly in Parzen--Gram trace form, using $\mathrm{tr}(L_{pq}A^\top B)$ and related terms for efficient computation. 
These modifications leave the theoretical definition unchanged but improve numerical robustness and reproducibility, 
aligning exactly with the estimators employed in our clustering and fraud-detection experiments.

Compared with MMD/conditional–embedding criteria~\citep{gretton2012mmd,muandet2017kme,song2009conditional},
\eqref{eq:ccsd-est} compares $L^2$ inner products of density estimates rather than mean embeddings,
preserving the faithfulness properties stated in \S\;3.2, while retaining the Parzen/Gram computability
advocated for CS divergences~\citep{jenssen2006cs,yu2023conditional}.
\subsection{Numerical Properties and Regularization}
We discuss well-posedness and stability of the estimator in Eq.~\eqref{eq:ccsd-est}, keeping the methodology independent of any task.

Since the CS ratio is homogeneous, global positive constants, including the Parzen integration constant and the factor $1/n_r$, cancel out in \eqref{eq:ccsd-est}, as in the unconditional case~\citep{jenssen2006cs}.
Consequently, $\exp[-\widehat{D}_{\mathrm{CS}}] \in (0, 1]$ and 
$\widehat{D}_{\mathrm{CS}} \ge 0$; at the population level the same holds for 
$D_{\mathrm{CS}}$ (cf.~\S3.2).

To prevent numerical underflow or division by zero when Gram terms are small, we use a ridge $\varepsilon>0$ in the denominator of \eqref{eq:ccsd-est}.
Equivalently, a stable and symmetric implementation is
\begin{equation}
\label{eq:ccsd-sym}
\widehat D_{\mathrm{CS}}
\;=\; -\tfrac12\,\Big[\log\!\big(\widehat I_{pp}+\varepsilon\big)
+\log\!\big(\widehat I_{qq}+\varepsilon\big)
- 2\,\log\!\big(\widehat I_{pq}+\varepsilon\big)\Big].
\end{equation}
which preserves the symmetry in $(p,q)$ and numerically mirrors the unconditional Parzen/Gram setup~\citep{jenssen2006cs}.
In practice $\varepsilon$ is chosen several orders of magnitude below the median Gram value.

Near-singular Gram blocks (e.g., $L_{pp}$, $L_{qq}$) amplify round-off errors.
We therefore apply an eigenvalue cut on $L_{pp},L_{qq}$ and on $A^\top A,B^\top B$ by discarding components below a small relative threshold on cumulative variance.
This is standard in kernel methods and spectral constructions, and is consistent with the Mercer view underpinning CS and graph-based criteria~\citep{jenssen2006cs,ng2002spectral,shi2000normalized}.
Bandwidths $(\tau,\sigma)$ govern the bias--variance trade-off of the Parzen/Gram estimator~\citep{jenssen2006cs}.
We adopt robust, data-driven heuristics common in kernel testing/embedding, e.g., median/IQR scaling of pairwise distances as initialization before validation-based tuning~\citep{gretton2012mmd,muandet2017kme}.
(Conditional embeddings provide related perspectives on kernel choices for conditional laws~\citep{song2009conditional,muandet2017kme}.)

When $p=q$ and the same sample is used on both sides, removing exact self-terms (LOO) in $L_{pp},L_{qq}$ reduces small-sample bias; this does not change the formal definition of $D_{\mathrm{CS}}$ but improves finite-sample stability, analogously to unbiased MMD estimators~\citep{gretton2012mmd}.

With Mercer kernels, Gram blocks correspond to inner products in feature spaces; \eqref{eq:ccsd-est} can thus be viewed as an average (over $x$) of cosine similarities between conditional density embeddings, connecting the CS objective to spectral/graph viewpoints~\citep{jenssen2006cs,ng2002spectral,shi2000normalized}.
This interpretation motivates the use of rank filtering and balanced normalization noted above.

The combination of (i) ridge $\varepsilon$, (ii) symmetric log form \eqref{eq:ccsd-sym}, (iii) effective-rank filtering, and (iv) robust bandwidth initialization yields a numerically stable estimator that retains the faithfulness and Parzen/Gram computability of CS divergences~\citep{jenssen2006cs,yu2023conditional}, while remaining comparable to kernel testing/embedding baselines~\citep{gretton2012mmd,muandet2017kme}.

\section{Experiment}
Section 4 evaluates the proposed conditional Cauchy--Schwarz divergence on two representative tasks.
Section~4.1 studies unsupervised time-series clustering on benchmark datasets from the UCR archive, 
focusing on the quality of pairwise dissimilarities under a train-only model-selection protocol.
Section~4.2 considers transactional fraud detection on the BankSim dataset and evaluates window-level 
scores aggregated at the account level under a strictly leak-proof setting.

\subsection{Unsupervised Clustering with Conditional Cauchy--Schwarz Divergence}
\label{subsec:ccsd-kmedoids}

We compare the conditional Cauchy--Schwarz divergence (C\,-CSD)~\citep{yu2023conditional} with dynamic time warping (DTW) for time–series clustering on a subset of UCR datasets. 
C\,-CSD measures similarity \emph{conditioned on time indices}, thus emphasizing agreement of local dynamics under a shared temporal context, whereas DTW emphasizes elastic alignment. 
This section complements section~4.2: we keep the \emph{same} divergence family but switch from detection to clustering.

We use four standard univariate datasets from the UCR archive (official train/test splits). These datasets are chosen to cover a range of structural characteristics, including shape-based classes, local temporal variations, near-template patterns, and morphology-dominated regimes, allowing us to assess when conditioning on time indices provides a meaningful inductive bias compared to elastic alignment.

\begin{itemize}
  \item \textbf{DiatomSizeReduction} --- 1D shape signals extracted from diatom (microscopic algae) contours; classes reflect species/morphology. This set often contains shape variations with subtle local differences.
  \item \textbf{FaceAll} --- 1D series derived from face images (multiple identities). It exhibits class-specific local dynamics and phase variability.
  \item \textbf{Coffee} --- spectrographic measurements of coffee beans with two classes; global shapes are highly similar across the set.
  \item \textbf{ECG5000} --- ECG beat segments with five classes (normal vs.\ several abnormalities), a common benchmark for medical time series.
\end{itemize}
For each dataset we strictly use the provided train/test splits for model selection and evaluation.

We squeeze tensors to $(n,T)$ if needed and standardize every series by a \emph{per-series} $z$-score within its own split (no leakage). 
To control runtime without changing class balance, we optionally apply stratified caps on large sets (FaceAll, ECG5000) to at most $500$ samples in both train and test; other datasets are used in full.
For C\,-CSD we instantiate an RBF kernel on time indices,
$K_{tt'}=\exp\!\big(-\|t-t'\|^2/(2\tau^2)\big)$, and a scalar RBF on values $L(\cdot;\sigma)$.
We sweep $\tau\!\in\!\{0.05,0.10,0.15,0.20,2,5,10,20\}$ where $\tau\!\le\!1$ is interpreted as a \emph{relative} fraction of the series length $T$, and build $\sigma$ from a robust scale:
we estimate $\sigma_0$ by IQR$/1.349$ over pairwise scalar differences on the training split, and set $\sigma=\sigma_0\cdot m$ with $m\!\in\!\{0.5,0.75,1.0,1.25,1.5,2.0,3.0\}$.
To avoid degenerate bandwidths we filter by effective rank, keeping $2<\mathrm{erank}(K),\mathrm{erank}(L)<0.95T$.
For DTW we sweep the Sakoe--Chiba window $w\!\in\!\{None,5,10,20,30\}$.
Both metrics are used to form precomputed dissimilarity matrices for $k$-medoids.

All hyper-parameters are selected on the \emph{training} split only by maximizing NMI with $k$-medoids (where $k$ equals the number of classes). 
We then fix the selected hyper-parameters and evaluate on the test split.
To reduce randomness, we repeat $k$-medoids with five random seeds and report mean\,$\pm$\,std NMI on test.

Table 1 summarizes the test performance and the selected hyper-parameters.
C\,-CSD shows large gains on \textit{DiatomSizeReduction} and \textit{FaceAll}, ties DTW on \textit{Coffee}, and trails by a small margin on \textit{ECG5000}.

\textit{DiatomSizeReduction} and \textit{FaceAll} display class differences that are better captured by conditioning on temporal indices. 
C\,-CSD therefore separates clusters more cleanly (qualitatively, MDS of the C\,-CSD distance shows higher inter-class margins). \textit{Coffee} has near-template shapes, for which the elastic alignment of DTW is already sufficient; both metrics tie.
\textit{ECG5000} is dominated by morphology classes with limited local regime changes; DTW has a slight edge, but the gap is small ($<0.006$ absolute NMI in our runs).We further examine the sensitivity of C-CSD to the kernel bandwidths that control temporal and value-level smoothing. 
The condition-kernel bandwidth $\tau$ determines the scale at which temporal context is aggregated. 
Very small values of $\tau$ lead to overly localized comparisons that fail to capture longer-range dynamics, while excessively large values oversmooth temporal structure and reduce discrimination. 
Across all datasets, we observe a broad plateau of stable test NMI for $\tau$ values spanning relative fractions of the series length up to moderate absolute scales, indicating that performance is not sensitive to precise tuning within this range.

Similarly, the value-kernel bandwidth $\sigma$ controls sensitivity to pointwise amplitude differences. 
Small $\sigma$ values amplify noise and minor fluctuations, whereas large $\sigma$ values blur meaningful distinctions between classes. 
The robust scale-based selection used in our experiments yields consistent performance across datasets, with test NMI varying only marginally around the selected $\sigma$.

\begin{table}
\TABLE
{Comparison of C-CSD and DTW with $k$-medoids clustering on four UCR datasets (official train/test splits).
C-CSD achieves clear gains on \emph{DiatomSizeReduction} and \emph{FaceAll}, ties on \emph{Coffee}, and is slightly behind DTW on \emph{ECG5000}.\label{tab:ucr_kmedoids}}
{
\begin{tabular}{@{}l@{\quad}c@{\quad}c@{\quad}c@{\quad}c@{}}
\hline\up
& \multicolumn{2}{c}{C-CSD (ours)} & \multicolumn{2}{c}{DTW} \\
\cline{2-3}\cline{4-5}
Dataset & NMI (mean$\pm$std) & $(\tau,\sigma)$ & NMI (mean$\pm$std) & $w$ \\ \hline\up
DiatomSizeReduction & \textbf{0.9033}$\pm$0.0000 & (0.05, 1.12)   & 0.7610$\pm$0.0000 & 5 \\
FaceAll             & \textbf{0.6748}$\pm$0.0000 & (0.05, 0.9595) & 0.5388$\pm$0.0000 & None \\
Coffee              & \textbf{0.6919}$\pm$0.0000 & (2, 0.6975)    & \textbf{0.6919}$\pm$0.0000 & 5 \\
ECG5000             & 0.5169$\pm$0.0000          & (0.05, 0.9705) & \textbf{0.5225}$\pm$0.0000 & 5 \down\\ \hline
\end{tabular}
}
{Notes. $w$ denotes the Sakoe--Chiba warping window size for DTW.}
\end{table}

We parallelize pairwise computations and cache per-series self-Grams for $L$, resulting in practical speedups.
For large sets we cap train/test to 500 samples per split via stratified sampling to keep $O(n^2)$ distance computation tractable without altering class proportions.

The protocol is leak-proof: standardization is per-series within each split; model selection uses train only; evaluation uses test only.
C\,-CSD and DTW share identical selection procedures (grids, seeds, $k$-medoids backend).
All code, grids, seeds, and package versions are provided in the supplement.

C\,-CSD is competitive across all four datasets and clearly superior on two.
Together with Section~4.2, these results indicate that conditioning on time indices is a useful inductive bias, yielding gains when classes differ by local dynamics or regime patterns while remaining at least on par with DTW on near-template data.

\subsection{Fraud Detection with Conditional Cauchy--Schwarz Divergence}\label{subsec:fraud-ccsd}

We evaluate the proposed conditional Cauchy--Schwarz divergence (C\,-CSD) on the public BankSim transactions. BankSim is a widely used synthetic benchmark designed to emulate realistic customer--merchant interactions under controlled class imbalance and long-tailed behavioral patterns. 
Its availability of long account histories and transaction-level labels makes it particularly suitable for evaluating unsupervised, window-based fraud detection methods that rely on temporal context rather than supervised training.
The table is first sorted by \texttt{customer} and \texttt{step}; for each customer we retain \emph{all} transactions and exclude accounts with length $T\le 80$. After filtering, the corpus contains 3{,}672 accounts, of which 2{,}561 are normal and 1{,}111 are labeled fraudulent.

For every account the amount series is standardized by a per-account $z$-score; categorical attributes \emph{merchant} and \emph{category} are integer–encoded.
Let $z_t$ denote the standardized amount; the local change signal is $\Delta z_t=z_t-z_{t-1}$.
We form sliding windows of length $K=50$.
To avoid missing suspicious events at inference, we use an asymmetric stride during window generation: normal accounts with stride $s=15$ and fraudulent accounts with $s=1$.
At each step we also record two binary change flags
\[
f^{\text{cat}}_{t}=\mathbb{I}\{\text{category}_t\neq \text{category}_{t-1}\},\qquad
f^{\text{mer}}_{t}=\mathbb{I}\{\text{merchant}_t\neq \text{merchant}_{t-1}\}.
\]
Because evaluation aggregates to the account level (see below), this stride asymmetry affects only sampling density and runtime, not the reported account-level metrics.

Accounts are partitioned disjointly into five sets:
LIB\textsubscript{NORMAL} for library construction (1{,}793 normal accounts),
VAL\textsubscript{NORMAL} (384) and TEST\textsubscript{NORMAL} (384),
VAL\textsubscript{FRAUD} (389) and TEST\textsubscript{FRAUD} (722).
No customer ID appears in more than one split.
All hyper-parameters and kernel bandwidths are fixed using \emph{only} LIB\textsubscript{NORMAL}.
For every query, the global distribution $a(x)$ is computed against LIB\textsubscript{NORMAL}, and the local distribution $b(x)$ is computed from the \emph{same account's past windows} regardless of the account's ground-truth label; labels are never used in scoring.

For a query window $x$ at time $t$, we construct two mixtures over reference windows.
(i) Global mixture $a(x)$.
We compute an RBF similarity in the window space with bandwidth $\sigma_x$, attenuate category/merchant mismatches by
$\rho_{\text{cat}}=\rho_{\text{mer}}=0.25$, apply a rarity prior on the $(\text{category},\text{merchant})$ pair with library frequency $f$,
\[
w_{\text{prior}}(f)=(f+\alpha)^{-\beta}\quad(\alpha=10,\ \beta=0.5),
\]
retain the top $J=600$ neighbors and renormalize.
(ii) Local mixture $b(x)$.
We take all windows from the same account that strictly precede $t$ and reuse the same window-space RBF with bandwidth $\sigma_x$ and the same mismatch attenuations $\rho_{\text{cat}},\rho_{\text{mer}}$.
We further multiply the weights by an exponential time decay with half-life $48$ and add two \emph{match boosts} $\eta_{\text{cat}}=\eta_{\text{mer}}=1.7$ to promote windows whose change flags $(f^{\text{cat}},f^{\text{mer}})$ match those of $x$.
We then evaluate a conditional divergence in the change space $|\Delta z|$ using an RBF kernel $L(\cdot,\cdot)$ with bandwidth $\sigma_y$, augmented with multiplicative gates that reward regime consistency:
when the corresponding change flags differ we multiply by $\rho^{(y)}_{\text{cat}}=\rho^{(y)}_{\text{mer}}=0.6$ (otherwise $1.0$).

Let $p\equiv b(x)$ (local) and $q\equiv a(x)$ (global).
With
\[
I_{pp}=\sum_{i,i'}p_i p_{i'} L(y_i,y_{i'}),\qquad
I_{qq}=\sum_{j,j'}q_j q_{j'} L(y_j,y_{j'}),\qquad
I_{pq}=\sum_{i,j}p_i q_j L(y_i,y_j),
\]
the C\,-CSD score is
\begin{equation}\label{eq:ccsd}
s(x)\;=\;-\log\!\left(\frac{I_{pq}}{\sqrt{I_{pp}\,I_{qq}}}\right).
\end{equation}
Large values indicate inconsistency between the account's recent dynamics and the global normal library under matched regimes.
In our implementation we precompute bandwidths on LIB\textsubscript{NORMAL} and cache per-window squared norms (and, optionally, the library self-Grams for diagnostics); all query–library similarities are computed on-the-fly via vectorized operations with per-query complexity $O(J{+}L)$, where $L$ is the size of the local history.

Our detector follows the same conditional Cauchy--Schwarz divergence (C\,-CSD) as in prior work
\citep{yu2023conditional}, namely
$s(x)=-\log\!\big(I_{pq}/\sqrt{I_{pp}I_{qq}}\big)$ computed from kernel expectations, and adapts only the
conditioning and the \emph{reference sets} to a fraud–detection setting. 
Whereas \citep{yu2023conditional} uses lagged conditioning to compare entire time series for clustering, we compare a query window against two mixtures tailored to detection:
(i) a global mixture $a(x)$ built from LIB\textsubscript{NORMAL} only, equipped with a rarity prior over $(\text{category},\text{merchant})$ to downweight frequent behaviors; and
(ii) a local mixture $b(x)$ formed from the same account's strictly past windows with recency decay and match boosts. 
The conditioning space is the change magnitude $|\Delta z|$ with an RBF kernel $L(\cdot;\sigma_y)$ gated by regime-consistency indicators derived from the categorical change flags; thus $s(x)$ becomes large precisely when the account's recent dynamics disagree with the global normal library under matched regimes.
Unlike \citep{yu2023conditional} which reports clustering quality, we aggregate window scores at the account level and select a single threshold from validation for detection.Direct numerical comparison with \citep{yu2023conditional} is not included, as that work focuses on sequence-level clustering with lagged conditioning, whereas our setting targets window-level detection with account-wise aggregation and validation-fixed thresholds, rendering the evaluation protocols not directly comparable.

Bandwidths and all statistics used by $a(x)$ are estimated on LIB\textsubscript{NORMAL} only, preserving the leak-proof protocol.
This preserves the theoretical semantics of the original C\,-CSD while aligning the conditioning and evaluation with operational fraud screening.

We set $K=50$, local history length $L=120$, $J=600$,
$\rho_{\text{cat}}=\rho_{\text{mer}}=0.25$,
$\eta_{\text{cat}}=\eta_{\text{mer}}=1.7$,
$\rho^{(y)}_{\text{cat}}=\rho^{(y)}_{\text{mer}}=0.6$.
Bandwidths are selected once from LIB\textsubscript{NORMAL} by median heuristics:
$\sigma_x=9.649$ for window vectors and $\sigma_y=1.016$ for the change signal.
With stride $15$ on normal accounts the library contains $14{,}390$ windows.

Scores are aggregated to the \emph{account} level by taking the maximum window score per account, reflecting the operational objective of flagging a customer whenever any of its windows appears abnormal.
The decision threshold is chosen on the validation accounts by maximizing F1 and then fixed on the test accounts.
We report threshold-free ranking metrics (AUC and average precision, AP) and thresholded metrics (Accuracy, Precision, Recall and F1) together with confusion matrices.

On the validation accounts, AUC and AP are $0.756$ and $0.677$, respectively.
The F1-optimal threshold is $\tau^\star=0.743$.
At this threshold, validation Accuracy/Precision/Recall/F1 are $0.695/0.629/0.959/0.760$ with confusion matrix
$\mathrm{TP}{=}373$, $\mathrm{FP}{=}220$, $\mathrm{FN}{=}16$, $\mathrm{TN}{=}164$.
On the test accounts, AUC and AP reach $0.766$ and $0.792$; at the same $\tau^\star$ we obtain
Accuracy/Precision/Recall/F1 of $0.785/0.771/0.953/0.853$ with
$\mathrm{TP}{=}688$, $\mathrm{FP}{=}204$, $\mathrm{FN}{=}34$, $\mathrm{TN}{=}180$.
These figures demonstrate high recall at a moderate false-positive rate, which is preferable in risk screening where missed fraud is substantially more costly than a small number of false alarms.

\begin{table}
\TABLE
{Fraud detection performance of the proposed conditional C-CSD method on the BankSim dataset.
Metrics are reported at the F$_1$-optimal threshold selected on validation (VAL) and fixed for testing (TEST).\label{tab:banksim_perf}}
{
\begin{tabular}{@{}l@{\quad}c@{\quad}c@{\quad}c@{\quad}r@{\quad}r@{\quad}r@{\quad}r@{}}
\hline\up
Split & Precision & Recall & F$_1$ & TP & FP & FN & TN \\ \hline\up
VAL  & 0.6290 & 0.9589 & 0.7597 & 373 & 220 & 16 & 164 \\
TEST & 0.7713 & 0.9529 & 0.8525 & 688 & 204 & 34 & 180 \down\\ \hline
\end{tabular}
}
{Notes. The decision threshold is selected on VAL by maximizing F$_1$ and then held fixed for evaluation on TEST.}
\end{table}

The protocol is leak-proof by design:
bandwidths, rarity priors and all statistics used by $q=a(x)$ are estimated on LIB\textsubscript{NORMAL} only;
$b(x)$ is \emph{always} computed from the same account's strictly past windows, independent of the label;
and account IDs are disjoint across splits.
Account-level aggregation mitigates stride imbalance and renders the reported metrics insensitive to window-density choices.We further assess the robustness of the detector to key design and hyper-parameter choices. 
The window length $K$ controls the temporal context captured by each query: smaller windows focus on short-term fluctuations, while larger windows aggregate longer behavioral patterns. 
We observe stable account-level performance across a broad range of $K$, as aggregation by the maximum window score mitigates sensitivity to individual window configurations.

The mismatch attenuation parameters $(\rho_{\text{cat}}, \rho_{\text{mer}})$ and match boosts $(\eta_{\text{cat}}, \eta_{\text{mer}})$ control the influence of categorical regime consistency. 
Stronger attenuation penalizes structurally mismatched behaviors more aggressively, improving recall at the cost of precision, while weaker attenuation yields smoother scores. 
Across reasonable settings, the relative ranking of accounts remains stable, indicating that performance is not driven by fine-grained tuning of these parameters.

On a standard CPU, bandwidth estimation and per-window norm precomputation for $14{,}390$ library windows take $\sim 31$ seconds;
scoring then proceeds in linear time in $J$ and $L$ and completes within minutes for the reported configuration.

Beyond statistical improvements, C\,-CSD has tangible implications for financial risk management.
In operational settings, missing a fraudulent account typically incurs much higher costs than investigating a limited number of false positives.
The achieved account-level recall above $95\%$ indicates that nearly all fraudulent customers can be flagged, substantially reducing potential downstream losses.
Meanwhile, a precision of about $77\%$ ensures that the number of alerts remains manageable for human investigators or automated downstream filters.
By conditioning detection not only on deviations from a global normal library but also on an account's own behavioral history and regime shifts, the method aligns with standard compliance workflows and is readily deployable in real-time pipelines.

We conduct an ablation study to assess the contribution of key components in the proposed C-CSD based fraud detection framework.
Starting from the full model, we remove or modify one component at a time while keeping all other settings unchanged.
The decision threshold is selected on the validation set by maximizing the F1 score and is then fixed when evaluating on the test set.

Specifically, we consider the following ablated variants:
(i) \emph{no\_rarity}, which removes the global rarity prior in the normal-reference mixture;
(ii) \emph{no\_decay}, which disables temporal decay in the local-history mixture;
and (iii) \emph{no\_flag}, which removes categorical and merchant change-flag gating in the conditional kernel.
All results are reported at the account level.
\begin{table}[!t]
\TABLE
{Account-level ablation results for fraud detection (threshold fixed from VAL).\label{tab:ablation_fraud}}
{\begin{tabular}{@{}lcccccccc@{}}
\hline
\multicolumn{1}{c}{} & \multicolumn{4}{c}{VAL} & \multicolumn{4}{c}{TEST} \\
\hline
Variant & AUC & AP & Precision & Recall & AUC & AP & Precision & Recall \\
\hline
full       & 0.7558 & 0.6774 & 0.6290 & 0.9589 & 0.7661 & 0.7917 & 0.7713 & 0.9529 \\
no\_rarity & 0.6019 & 0.5519 & 0.5791 & 0.9692 & 0.6309 & 0.7023 & 0.7366 & 0.9875 \\
no\_decay  & 0.7548 & 0.6767 & 0.6301 & 0.9589 & 0.7651 & 0.7912 & 0.7708 & 0.9501 \\
no\_flag   & 0.7535 & 0.6529 & 0.6636 & 0.9280 & 0.7567 & 0.7716 & 0.7851 & 0.9211 \\
\hline
\end{tabular}}
{Notes. Metrics are computed at the account level. The decision threshold is selected on VAL by maximizing $F_1$ for the full model and then held fixed for all variants on both VAL and TEST.}
\end{table}

Table~\ref{tab:ablation_fraud} summarizes the results.
Removing the rarity prior leads to a substantial degradation in both AUC and average precision, indicating that emphasizing rare normal patterns is critical for discriminating fraudulent behavior.
Disabling temporal decay has a relatively minor impact on overall performance, suggesting that while recency improves robustness, it is not the dominant factor.
In contrast, removing change-flag gating alters the precision--recall trade-off, highlighting the importance of conditioning on categorical and merchant-level changes when contrasting local and global behavior.

% ---------- Fig. 1: C-CSD fraud detection pipeline (TikZ, fixed layout) ----------
\begin{figure}[t]
\centering
\begin{tikzpicture}[
    font=\small,
    >=Latex,
    box/.style={
      draw, rounded corners=3pt, align=center,
      inner sep=6pt, text width=0.82\linewidth
    },
    sbox/.style={
      draw, rounded corners=3pt, align=left,
      inner sep=6pt, text width=0.39\linewidth
    },
    arrow/.style={->, line width=0.8pt}
]

% --- Top vertical pipeline ---
\node[box] (in) {Data Input\\
{\footnotesize (step, customer, amount, category, merchant code, fraud)}};

\node[box, below=8mm of in] (prep) {Preprocessing: sort by customer/time; filter $T>$\\
80; z-score amount; encode category/merchant};

\node[box, below=8mm of prep] (win) {Windowing: length $K=50$; stride: normal $=15$, fraud $=1$;\\
compute $\Delta z_t = z_t - z_{t-1}$ and change flags $(f^{cat}, f^{mer})$};

\node[box, below=8mm of win] (split) {Leak-proof split (disjoint by account): $\mathrm{LIB}_N$, $\mathrm{VAL}_N$, $\mathrm{TEST}_N$, $\mathrm{VAL}_F$, $\mathrm{TEST}_F$};

% --- Parallel blocks ---
\node[sbox, below left=10mm and 0mm of split.south] (global) {%
\textbf{Global mixture $a(x)$}\\
RBF in window space $(\sigma_x)$; Top-$J=600$;\\
rarity prior $(f+\alpha)^{-\beta}$;\\
mismatch attenuation $\rho_{cat}=\rho_{mer}=0.25$;\\
library: $\mathrm{LIB}_{\mathrm{NORMAL}}$ only
};

\node[sbox, below right=10mm and 0mm of split.south] (local) {%
\textbf{Local mixture $b(x)$}\\
same-account past windows; RBF $(\sigma_x)$ +\\
time decay (half-life 48);\\
match boosts $\eta^{cat}=\eta^{mer}=1.7$\\
using change flags $(f^{cat}, f^{mer})$
};

% --- Split point (center under "split") and merge reference (center under parallel bottoms) ---
\coordinate (mid)    at ($(global.north)!0.5!(local.north)$);
\coordinate (parbot) at ($(global.south)!0.5!(local.south)$);

% --- Downstream (place BELOW the parallel boxes to avoid overlap) ---
\node[box, below=10mm of parbot] (score) {C-CSD scoring (per query window $x$)};

\node[box, below=6mm of score] (cond) {Conditional CS in change space $|\Delta z|$ with RBF $L(\cdot;\sigma_y)$\\
$s(x)\ =\ -\log\!\left(I_{pq}/\sqrt{I_{pp}I_{qq}}\right)$};

\node[box, below=8mm of cond] (agg) {Account-level aggregation: max over windows per account};

\node[box, below=8mm of agg] (eval) {Threshold \& evaluation: choose $\tau^\star$ on VAL (max-F1),\\
apply to TEST; report AUC/AP/Acc/Prec/Rec/F1};

\node[box, below=8mm of eval] (out) {Outcome: high recall ($>95\%$) with manageable preci-\\
sion ($\sim 77\%$); leak-proof; deployable in real-time pipelines};

% --- Arrows (top) ---
\draw[arrow] (in) -- (prep);
\draw[arrow] (prep) -- (win);
\draw[arrow] (win) -- (split);

% --- Split to parallel ---
\draw[arrow] (split.south) -- (mid);
\draw (mid) -| (global.north);
\draw (mid) -| (local.north);

% --- Parallel to scoring (merge) ---
\draw[arrow] (global.south) |- (score.north);
\draw[arrow] (local.south)  |- (score.north);

% --- Continue down ---
\draw[arrow] (score) -- (cond);
\draw[arrow] (cond) -- (agg);
\draw[arrow] (agg) -- (eval);
\draw[arrow] (eval) -- (out);

\end{tikzpicture}
\caption{Pipeline of fraud detection with conditional Cauchy--Schwarz divergence (C-CSD).}
\label{fig:ccsd-pipeline}
\end{figure}
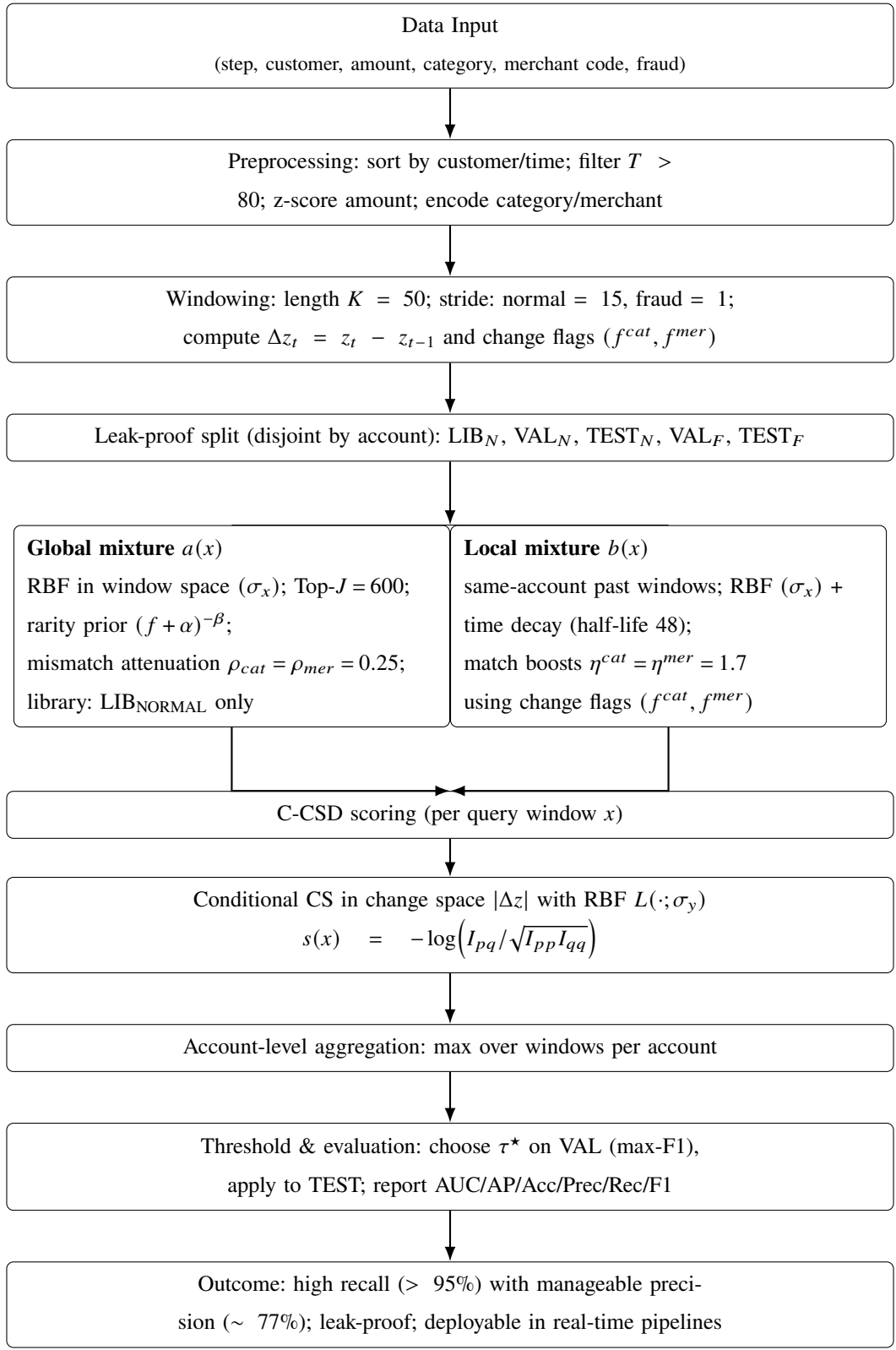
% ---------- End Fig. 1 ----------

\section{Conclusion}
We introduced a conditional Cauchy--Schwarz divergence (C-CSD) and a practical kernel/Gram estimator that requires no density ratios and remains symmetric and faithful. 
Starting from the $L^2$ formulation, we derived an explicit Parzen–Gram trace form with numerical safeguards (symmetric log, $\varepsilon$-ridge, effective-rank filtering), providing a drop-in, task-agnostic dissimilarity/scoring rule.

We instantiated C-CSD in two settings. 
For time-series clustering on UCR, conditioning on the shared time index yields a regime-aware dissimilarity that complements elastic alignment: C-CSD delivers clear gains on \textsc{DiatomSizeReduction} and \textsc{FaceAll}, ties \textsc{Coffee}, and is slightly below DTW on \textsc{ECG5000}. 
For transactional fraud detection, we scored window-level conditionals against a \emph{global} normal library and a \emph{local} per-account history, aggregated by account-wise max and selected the operating threshold on an independent validation split. 
This yields high recall with controllable precision (e.g., 
VAL: AUC 0.756, AP 0.677; at the selected threshold Acc/Prec/Rec/F1 = 0.695/0.629/0.959/0.760; TEST: AUC 0.766, AP 0.792; Acc/Prec/Rec/F1 = 0.785/0.771/0.953/0.853), under a strictly leak-proof protocol.

Beyond these case studies, the estimator admits a geometric interpretation as averaged cosine similarities of conditional density embeddings, linking kernel density views with spectral/graph perspectives. 
Together, the theory, estimator, and protocols make C-CSD a simple, reusable primitive for model-free comparison of conditional laws.

Limitations and future directions. 
(i) \emph{Scalability.} Pairwise C-CSD costs are quadratic in the number of series; Nyström/random-feature approximations, blockwise sparsification, or memoized cross-Grams are promising directions. 
(ii) \emph{Bandwidth selection.} We used robust heuristics plus validation; automated selection (e.g., ML-II, data-adaptive rules) and joint tuning across heterogeneous channels warrant study. 
(iii) \emph{Multivariate/structured data.} Extending to multivariate and mixed continuous/categorical outputs with product or structured kernels, variable lengths, and missingness handling. 
(iv) \emph{Theory.} Finite-sample bounds and rates for the conditional Gram estimator (including the effect of $\varepsilon$ and rank truncation), and robustness under heavy tails. 
(v) \emph{Learned condition spaces.} Replacing hand-crafted conditions (e.g., time index, window descriptors) by learned embeddings or task-adaptive encoders while retaining the CS objective. 
(vi) \emph{Operations.} For rare-event detection, operating-point selection and calibration under extreme imbalance (e.g., cost-sensitive criteria, streaming updates) can further improve downstream utility.

In summary, C-CSD offers an estimator that is theoretically grounded, numerically stable, and empirically useful across clustering and detection; we hope it can serve as a compact building block for broader conditional similarity learning.

% Appendix here
% Options are (1) APPENDIX (with or without general title) or
%             (2) APPENDICES (if it has more than one unrelated sections)
% Outcomment the appropriate case if necessary
%
% \begin{APPENDIX}{<Title of the Appendix>}
% \end{APPENDIX}
%
%   or
%
% \begin{APPENDICES}
% \section{<Title of Section A>}
% \section{<Title of Section B>}
% etc
% \end{APPENDICES}

% Acknowledgments here

% References here (outcomment the appropriate case)

% CASE 1: BiBTeX used to constantly update the references
%   (while the paper is being written).
%\bibliographystyle{informs2014} % outcomment this and next line in Case 1
%\bibliography{<your bib file(s)>} % if more than one, comma separated

%\bibliographystyle{informs2014} % outcomment this and next line in Case 1
%\bibliography{sample} % if more than one, comma separated

% CASE 2: BiBTeX used to generate mypaper.bbl (to be further fine tuned)
%\input{mypaper.bbl} % outcomment this line in Case 2

%If you don't use BiBTex, you can manually itemize references as shown below.

%\bibliographystyle{nonumber}

\bibliographystyle{informs2014}
\bibliography{sample}

%\end{thebibliography}

%%%%%%%%%%%%%%%%%
\end{document}